\documentclass[modern]{aastex62}
\usepackage{amsmath}

\shorttitle{spectrophotometric parallax}
\shortauthors{hogg, eilers, rix}

\newcommand{\documentname}{\textsl{Article}}
\newcommand{\sectionname}{Section}
\newcommand{\equationname}{equation}
\newcommand{\code}[1]{\texttt{\detokenize{#1}}}
\newcommand{\foreign}[1]{\textsl{#1}}
\newcommand{\etal}{\foreign{et al.}}
\newcommand{\eg}{\foreign{e.g.}}
\newcommand{\acronym}[1]{{\small{#1}}}
\newcommand{\project}[1]{\textsl{#1}}
\newcommand{\apogee}{\project{\acronym{APOGEE}}}
\newcommand{\gaia}{\project{Gaia}}
\newcommand{\wise}{\project{\acronym{WISE}}}
\newcommand{\zmass}{\project{\acronym{2MASS}}}
\newcommand{\sdssiv}{\project{\acronym{SDSS-IV}}}
\newcommand{\sdssv}{\project{\acronym{SDSS-V}}}

\newcommand{\T}{^{\mathsf{T}}}
\DeclareMathOperator*{\argmin}{argmin}
\newcommand{\logg}{\log g}
\newcommand{\BP}{{G_\mathrm{BP}}}
\newcommand{\RP}{{G_\mathrm{RP}}}
\newcommand{\gparallax}{{\varpi^{(\mathrm{a})}}}
\newcommand{\sparallax}{{\varpi^{(\mathrm{sp})}}}
\newcommand{\gsigma}{{\sigma^{(\mathrm{a})}}}
\newcommand{\ssigma}{{\sigma^{(\mathrm{sp})}}}

\begin{document}\sloppy\sloppypar\raggedbottom\frenchspacing 

\title{\textbf{%
Spectrophotometric parallaxes with linear models:\\
Accurate distances for luminous red-giant stars
}}

\author[0000-0003-2866-9403]{David W. Hogg}
\affiliation{Center for Cosmology and Particle Physics, Department of Physics, New York University, 726 Broadway, New York, NY 10003, USA}
\affiliation{Center for Data Science, New York University, 60 Fifth Ave, New York, NY 10011}
\affiliation{Max-Planck-Institut f\"ur Astronomie, K\"onigstuhl 17, 69117 Heidelberg, Germany}
\affiliation{Flatiron Institute, 162 Fifth Ave, New York, NY 10010, USA}

\author[0000-0003-2895-6218]{Anna-Christina Eilers}
\affiliation{Max-Planck-Institut f\"ur Astronomie, K\"onigstuhl 17, 69117 Heidelberg, Germany}
\affiliation{International Max Planck Research School for Astronomy \& Cosmic Physics at the University of Heidelberg}

\author[0000-0003-4996-9069]{Hans-Walter Rix}
\affiliation{Max-Planck-Institut f\"ur Astronomie, K\"onigstuhl 17, 69117 Heidelberg, Germany}

\begin{abstract}\noindent
With contemporary infrared spectroscopic surveys like \apogee,
red-giant stars can be observed to distances and extinctions
at which \gaia\ parallaxes are not highly informative.
Yet the combination of effective temperature, surface gravity, composition, and age---all
accessible through spectroscopy---determines a giant's luminosity.
Therefore spectroscopy plus photometry should enable 
precise spectrophotometric distance estimates.
Here we use the \apogee--\gaia--\zmass--\wise\ overlap to train a data-driven model
to predict parallaxes for red-giant branch stars with $0<\logg\leq2.2$ (more luminous
than the red clump).
We employ (the exponentiation of)
a linear function of \apogee\ spectral pixel intensities and multi-band photometry
to predict parallax spectrophotometrically.
The model training involves no logarithms or inverses of the \gaia\ parallaxes,
and needs no cut on the \gaia\ parallax signal-to-noise ratio.
It includes an L1 regularization to zero out the contributions of
uninformative pixels.
The training is performed
with leave-out subsamples such that no star's astrometry is used even indirectly in its
spectrophotometric parallax estimate.
The model implicitly performs a reddening and extinction correction in its parallax prediction,
without any explicit dust model.
We assign to each star in the sample a new spectrophotometric parallax estimate;
these parallaxes have uncertainties of a few to 15 percent, depending on data quality,
which is more precise than the \gaia\ parallax for the vast majority of targets, and
certainly any stars more than a few kpc distance.
We obtain 10-percent distance estimates out to heliocentric distances of 20\,kpc,
and make global maps of the Milky Way's disk.
\end{abstract}


\section*{~}\clearpage
\section{Introduction} \label{sec:intro}

If we want to make precise kinematic and element-abundance maps
of the Milky Way disk out to large heliocentric radii,
and especially through the extinction to the other side of the Galactic Center, we will
need to make use of luminous red giants, and we will need to observe in
the infrared.
These arguments motivate the \apogee\ (\citealt{apogee}),
\apogee-2, and \sdssv\ (\citealt{sdssv})
projects, which take spectra in the infrared, and deliver detailed abundances
along the entire red-giant branch up to the tip.
The maps made with these data will reveal the dynamics of the Milky Way and its disk,
and show us how and where stars form, and how they migrate around the
disk with time.

As we operate these spectroscopic surveys,
the \gaia\ Mission (\citealt{gaia}) has also been revolutionizing our view of the disk.
It delivers good proper motions over most of the Galaxy,
and extremely valuable parallax information locally.
However, \gaia\ does not deliver precise geometric parallaxes over a large fraction of
the disk, and certainly not in the dusty and crowded regions.
For these reasons, the value of \gaia\ in these infrared spectroscopy projects is not
to deliver distance information directly, but rather to calibrate stellar models,
which then deliver distance information through relationships between stellar
luminosities and spectral characteristics.

Luminous red-giant stars are valuable for mapping the Milky Way disk for a number
of reasons, only one of which is their great luminosities (and hence brightnesses
even at large distance).
They are very common stars, much more frequent than distance-indicating
stars like Cepheids.
They are produced by stellar populations at all metallicities and almost all ages,
so they can be found in all parts of the Galaxy.
They have temperatures and surface gravities such that it is
straightforward to spectroscopically measure metallicities,
detailed chemical abundances, and stellar parameters, even including ages (\citealt{martig, nessage}).
The red-giant branch is photometrically near-orthogonal to reddening vectors by dust,
so they are easy to de-redden or dust-correct.
And finally, in physical models of red giants, their predicted luminosities are simple
functions of composition, surface gravity, temperature, and age.
Thus, if we can measure these properties of red giants, we can (in principle) predict
their luminosities and use them as standard candles.
That is, red giants are standardizable candles.

In this project, we develop purely data-driven techniques to predict
red-giant parallaxes or distances with spectroscopy and photometry.
Our method is predicated on the expectation that red-giant stars are dust-correctable,
standardizable candles with such data,
but it makes no use whatsoever of any stellar interior,
stellar exterior, or dust-reddening models.
We only use these physical expectations about red giants to recommend
data features---inputs---for a completely data-driven model.
The data-driven method uses patterns in the observed data themselves
to discover the relationships between spectral features in the spectra of the stars,
photometry (including colors), and parallax (or distance).

Data-driven models are very often more accurate than physical models of stars
for this kind of work, because they have the flexibility to discover
regularities or offsets in the data that are not adequately captured by the physical
models.
They do better than physical models when there are lots of training data
with properties that are good for performing regressions.

Simple data-driven models (like the ones we will employ here)
also suffer from many limitations for this kind of work:
Because they are constrained only by the data, and not physical law,
they can learn relationships that we know
to be physically unlikely, and (by the same token) they can't benefit from our
physical knowledge.
They are required, in some sense, to spend some of the information in the data on
learning physical laws or relationships.
Not all of the information in the data is being used
optimally when we ignore our physical models.
Relatedly, they are not (usually) interpretable or generalizable or useful for
extrapolations significantly outside the training set:
A data-driven model trained with one kind of data can not usually be used in another
context with very different input data.
And the internal properties of the model are not useful, in general, for improving
our understanding or intuition about the astrophysics in play.

All that said, our goal here is to produce a precise mapping tool for surveys
like \apogee\ and \sdssv.
So we accept the limitations of the data-driven models in exchange for their
performance.
This work builds on the success we have seen in building other kinds of
data-driven tools for  stellar spectroscopy, for example to measure stellar
parameters (\citealt{cannon}), abundances (\citealt{ho, casey, nessdopp}),
and masses and ages (\citealt{nessage}).

There is a huge range of complexity or capacity in data-driven
models, from linear regressions (what we use here) to the most advanced machine-learning
methods (like deep neural networks).
In this work we stay on the low-complexity side of this, as
we do not want to face the (literally combinatoric) choices in model structure that
advanced methods bring.
And we want to control or understand the smoothness or flexibility of the model.
Deep networks can model extremely complex function spaces, which is good in some
contexts, but bad when you know that the functional dependencies you hope to
model are smooth:
In that case, much of the information in the data can get spent on learning the
smoothness of the function.

What follows is a regression:
We use \gaia\ measurements of parallax to learn relationships between 
the spectroscopic and photometric properties of stars and their parallaxes.
We then use that learned model to estimate improved parallaxes for stars with
poor (or no) \gaia\ measurements.
This regression is very sensitive to certain aspects of the data or experimental
design.
One is that we will do far better with the regression if we have an accurate
model for noise process in the \gaia\ data.
For this reason, we make use of a justifiable likelihood function for the
\gaia\ astrometry (\citealt{gaialf}).
Another is that regressions can be strongly biased if we have a bad selection function
or data censoring.
This means that we cannot apply cuts to the \gaia\ astrometry or signal-to-noise
unless those cuts are correctly accounted for in our generative model for the
data sample.
Below, we will make no such cuts. That means that our training step has the odd
property that much of the training data is low in signal-to-noise; we include
training parallax measurements even if they are negative!
The fact that we are using a justified likelihood function ensures that these
data will only affect the model in sensible and justifiable ways.
The alternative---cutting to high signal-to-noise data---is the more traditional
approach, but it leads to biased inferences (unless the cuts are included correctly
in the likelihood function).
The method in which we make no cuts and keep an umodified, justified likelihood function
is unbiased, and simple.

\section{Assumptions of the method}\label{sec:assumptions}

Our position is that a methodological technique is correct inasmuch as
it delivers correct or best results \emph{under its particular assumptions}.
In that spirit, we present here the assumptions of the method
explicitly.
\begin{description}
\item[features]\footnote{Here and everywhere in this \documentname, we
  use the word ``feature'' to refer to an input or predictor or
  independent variable for the regression.
  That is, our parallax predictions will be linear combinations of features.
  This is not common terminology in astrophysics, but it is standard in the machine-learning literature.}
Perhaps our most fundamental assumption is that the parallax
of a star can be predicted from the features we provide, which are
the full set of (pseudo-continuum-normalized) pixels from the \apogee\ (\citealt{apogee}) spectrum,
plus the $G$, $\BP$, and $\RP$ photometry from \gaia\ (\citealt{gaia}),
plus the $J$, $H$, and $K_s$ photometry from \zmass\ (\citealt{zmass}),
plus the $W_1$ and $W_2$ photometry from \wise\ (\citealt{wise}).
That is, we assume that these spectrophotometric \emph{features} are sufficient
to predict the parallax in the face of variations in stellar
age, evolutionary phase, composition, and other parameters, and also interstellar
extinction.

\item[good features] We assume that the spectrophotometric features are known for
each star with such high fidelity (both precision and accuracy) that we do not
need to account for errors or uncertainties or biases in the features.
That is, we assume that the features are substantially higher in signal-to-noise than the
quantities we are trying to predict; in particular we are assuming that the photometry
and the spectroscopy is better measured than the astrometry.
This is true for most features for most stars, but it does not hold universally.

\item[representativeness] We assume that the training set constructed from the overlap
of \gaia\ and \apogee\ data sets constitutes a representative sample of stars,
sufficiently representative to train the model for all other stars.
Although this assumption is not terrible, it has a weakness:
The stars at greatest distances and greatest local (angular) stellar crowding have
the least precise \gaia\ parallax measurements, and therefore will effectively get
less weight in the fits performed to train the model.

\item[sparsity] We expect that only a small subset of the full complement of
\apogee\ spectral pixels will be relevant to the prediction of parallax
(the spectrophotometric parallax).
That is, we expect that many of the pixels will or ought to get no weight in the
final model that we use to predict luminosities and distances.

\item[linearity] Perhaps the most restrictive assumption of this work is that
the logarithmic parallax (or, equivalently, the distance modulus) 
can be predicted as a \emph{completely linear function} of
the chosen features. We are only assuming this linearity in a small range of stellar
surface gravity, but this assumption is strong, and limits strongly the
flexibility or capacity of the model.
We make it to ensure that our method is easy to optimize, and the results are easy
to interpret.
It is also the case that linear models extrapolate better than higher-order or more
flexible models; this is not extremely important to the present work, but it does
protect our predictions at the edges (in terms of temperature, surface gravity,
and chemical abundances) of our sample.

\item[likelihood] We assume that the \gaia\ parallaxes are generated by a particular
stochastic process, in which the difference between the \gaia-Catalog parallax (plus
a small offset, which we learn below) and the true parallax is effectively drawn from a
Gaussian with a width set by the \gaia-Catalog uncertainty on the parallax.
This is the standard assumption in all properly probabilistic \gaia-Mission inferences
to date, but it subsumes a number of related assumptions, like that the \gaia\ noise
model is correct, that the stars are only covariant with one another at negligible levels, and that
there are no significant outliers or large-scale systematics.
\end{description}

In addition to making the above assumptions, we also avoid various practices
with the data that are tempting but lead to important biases.
For example, we never cut on \gaia\ parallax or parallax signal-to-noise.
The common practices of cutting to parallaxes that are good to 20~percent
(see, for example, \citealt{trick, helmi}),
or Catalog parallaxes that are positive, or parallaxes that are smaller or
larger than something, are all practices that will bias results on stellar
collections.
That is, if you cut on parallax or parallax signal-to-noise and you subsequently
take an average of parallaxes for some population, or perform a regression (as we
do here), the results of that average or regression will be strongly biased.
We never cut on parallax or parallax signal-to-noise.
By using a justifiable likelihood function (the ``likelihood'' assumption above),
we can use all of the \gaia\ parallaxes without the low signal-to-noise and
negative parallaxes causing trouble for our regressions, and without the biases
that enter when cuts are made.

Along similar lines, we never assume that the distance is the inverse of the
measured parallax.
In what follows, a star's distance is a latent property of the star, which generates
the \gaia\ Catalog parallax through a noisy process (again, this is
the ``likelihood'' assumption above).
We never take the inverse or the logarithm of the measured parallax at any time.
This is related to the no-cuts point above:
If you take the inverse or logarithm of the parallax, you can't operate safely
on the negative and low signal-to-noise parallaxes, which in turn will require
making cuts on the sample, which will in turn bias the results.

Finally, we never use Lutz--Kelker corrections (\citealt{lk}) or distance
posteriors (\eg, \citealt{calj}). These both involve (implicit or explicit) priors on 
distance.
When multiple stars are combined as we combine them here (below),
use of distance posteriors instead of parallax likelihoods is not just
unjustified, but it also leads to an
effective raising of the distance prior to an enormous power.
That is, the effective prior on a $N$-star inference performed naively
with distance posteriors (made with a weak prior) can end up
bringing into the inference an \emph{exceedingly
strong prior}.
Therefore we don't use any prior-contaminated parallax or distance
inputs to the inference.

\section{Method}

There are not many choices
for building a model of the stars that is both justifiable probabilistically
and consistent with the assumptions stated in the previous \sectionname.
Here we lay out the model and methodology.
We apply the method to real data in the following {\sectionname s}.

The model for the parallax
and the log-likelihood function can be expressed heuristically as
\begin{eqnarray}
\gparallax_n &=& \exp(\theta\cdot x_n) + \mbox{noise}
\\
\chi^2(\theta) &\equiv& \sum_{n=1}^N \frac{[\gparallax_n - \exp(\theta\cdot x_n)]^2}{\gsigma_n^2}
\end{eqnarray}
where
$\gparallax_n$ is the \gaia\ measurement (adjusted; see below) of the astrometric parallax of star $n$,
the model is that the logarithm of the true parallax
can be expressed as a linear combination of the components
of some $D$-dimensional feature vector $x_n$,
$\theta$ is a $D$-vector of linear coefficients,
$\gsigma_n$ is the \gaia\ estimate of the uncertainty on the parallax measurement,
and $\chi^2(\theta)$ is (twice) the negative-log-likelihood for the parameters $\theta$
under the assumption of known Gaussian noise and
that there are $N$ independently measured stars $n$.
The feature vector $x_n$ contains photometry in a few bands, and all 7400-ish pixels
in the pseudo-continuum-normalized \apogee\ spectrum, so $D$ is on the order of 7400.

In addition, we assume that many entries in the parameter $D$-vector $\theta$ will be zero
(the ``sparsity'' assumption).
In order to represent this expectation,
we optimize not $\chi^2$ but a regularized objective function
\begin{eqnarray}
\hat{\theta} &\leftarrow& \argmin_{\theta}\left[\frac{1}{2}\,\chi^2(\theta) + \lambda\,||P\cdot\theta||_1^1\right]
\label{eq:objective}
\end{eqnarray}
where
$\lambda$ is a regularization parameter,
$P$ is a projection operator that selects out from the $\theta$ vector only those components
that pertain to the \apogee\ spectral pixels,
and $||y||_1^1$ is the L1-norm or sum of absolute values of the components of $y$.
This kind of regularization (L1) adds a convex term to the optimization and leads to
sparse optima.
The $P$ operator makes it such that we only regularize the components of $\theta$ that multiply
the spectral pixels; we ask for the spectral model to be sparse but we don't ask for the photometric
model to be sparse.
We set the value of $\lambda$ by cross-validation across the A/B split (see below).
In the end, the L1 regularization zeros out about 75 percent of the model coefficients.

This optimization is not convex---there are multiple optima in general.
In particular, there is a large, degenerate, pathological optimum where
the exponential function underflows, all parallaxes are predicted to vanish,
and there is no gradient of anything with respect to the parameter $D$-vector $\theta$.
This large, bad optimum (it is a local, not global, optimum) must be avoided in optimization.
In practice we avoid it by optimizing first for the very highest signal-to-noise
(better than 5-percent measurements of astrometric parallax)
Training Set stars, and then use that first optimum as a
good first guess or initialization for the full optimization.
This method works because in the limit of high signal-to-noise, the problem asymptotically approaches
the convex problem of L1-regularized linear least-square fitting.
Optimization is performed with \code{scipy.optimize} using the
\acronym{L-BFGS-B} algorithm (\citealt{lbfgsb}).

Once the model is optimized---and therefore $\hat{\theta}$ is determined---the
output of the model is a prediction of the parallax,
or really what we will call the \emph{spectrophotometric parallax}.
This spectrophotometric parallax
$\sparallax_m$ for any star $m$ in the validation or test set is
assigned according to
\begin{eqnarray}
\sparallax_m &\leftarrow& \exp(\hat{\theta}\cdot x_m)
\end{eqnarray}
where
$\hat{\theta}$ is the optimal parameter vector according
to \equationname~(\ref{eq:objective}),
and
$x_m$ is the feature $D$-vector for star $m$.
The model can be trained on a training set of stars and applied to
a validation set or a test set to make predictions.
The only requirements are that every test-set object $m$  must have a full feature
vector $x_m$ just as every training-set object $n$ must have a full feature
vector $x_n$.

The linear model permits straightforward propagation of uncertainty from the
feature inputs to the spectrophotometric parallaxes.
Although the model fundamentally assumes that the prediction features are noise-free
(see the ``good features'' assumption in \sectionname~\ref{sec:assumptions}),
there are in fact small uncertainties on the features that we can propagate to 
make uncertainty estimates on the spectrophotometric parallaxes.
\begin{eqnarray}
\ssigma_m^2 &\leftarrow& \sparallax_m^2\,\hat{\theta}^T\cdot C_m\cdot\hat{\theta}
\label{eq:unc}
\end{eqnarray}
where $\ssigma_m$ is the uncertainty on the inferred (or predicted)
spectrophotometric parallax $\sparallax_m$,
the right-hand side is a scalar product,
and $C_m$ is the covariance matrix of the input
features.
In what follows we don't actually instantiate the full covariance matrix; we presume
that input features are independently measured;
that is, we make $C_m$ is a diagonal matrix with the uncertainty variances of the elements
of $x_m$ along the diagonal.

We perform all of the fitting in a two-fold train-and-test framework,
in which the data are split into two disjoint subsets, A and B.
The model trained on the A data is used to predict or produce
the spectrophotometric parallaxes $\sparallax_m$ values (and hence
the distances and parallaxes) of the B data,
and the model trained on the B data is used to 
produce the $\sparallax_m$ values of the A data.
This ensures that, in the estimate of any individal star's
spectrophotometric parallax, none of the \gaia\ data pertaining to that star were used.
This makes the parallax estimates from the spectrophotometric feature vectors
$x_m$ statistically independent of the \gaia\ parallax measurements.
They are not independent globally---\gaia\ data were used to train the model---but
each star's spectrophotometric parallax estimate is independent
of the astrometric estimate from \gaia\ on a star-by-star basis.

\section{Data}

We take the \apogee\ (\citealt{aapogee, wapogee, apogee}) spectral data
from \sdssiv\ (\citealt{sdssiv}) \acronym{DR14} (\citealt{dr14}).
Because we want to make a purely linear model, which has very little capacity,
we restrict our consideration to a small region in stellar parameter space.
We cut the \apogee\ data down to the range $0<\logg<2.2$, which isolates
stars that are more luminous than the red clump.
The \apogee\ pipeline (\citealt{aspcap})
values of surface gravity $\logg$ (which we use) have uncertainties but
this cut leads to a clean sample of luminous red giants, and it is a cut
that is only on spectral properties of the stars (and not photometry nor astrometry).

From the \apogee\ data on these low-gravity stars, we take the spectral pixels,
of which there are 7405 (after cutting out pixels that rarely or never get data),
on a common rest-frame wavelength grid.
That is, every \apogee\ star is extracted on (or interpolated to)
the same wavelength scale.
In detail, we obtain the
the \apogee\ spectral pixels from the pipeline-generated \code{aspcapStar} files.
We then pseudo-continuum-normalize the spectra according to the procedure developed
in \project{The~Cannon} (\citealt{cannon}):
That is, the pseudo-continuum is a spline fit to a set
of pixels chosen to be insensitive to stellar parameters.
We use as our spectral data the normalized spectral pixel values.

Because the \apogee\ target selection is based on \zmass\ (\citealt{zmass}),
every \apogee\ star also comes with \zmass\ photometry.
That is, for each star $n$,
we have \zmass\ near-infrared photometry $J_n$, $H_n$, and $K_n$.

We used the \gaia\ (\citealt{gaia}) \acronym{DR2} (\citealt{gaiadr2}) Data Archive (\citealt{gaiaarchive})
official matches (\citealt{xmatch}) to match the \apogee+\zmass\ stars
to the \wise\ Catalog (\citealt{wise}),
according to the \gaia\ Archive internal match criteria.
This gives, for each matching star $n$,
mid-infrared photometry $W_{1n}$ and $W_{2n}$ at 3.6 and 4.5\,$\mu$m
In detail we use the \code{w1mpro} and \code{w2mpro} Catalog entries.

We match this full-match catalog to the \gaia\ \acronym{DR2}
using the \gaia\ Archive official match (\citealt{xmatch}) to the the \zmass\ IDs.
We take from the \gaia\ \acronym{DR2} Catalog the photometric data
$G_n$ (\code{phot_g_mean_mag}),
$\BP_n$ (\code{phot_bp_mean_mag}),
and $\RP_n$ (\code{phot_rp_mean_mag}),
which will become part of the feature vector $x_n$.

\begin{figure}
\includegraphics[width=\textwidth]{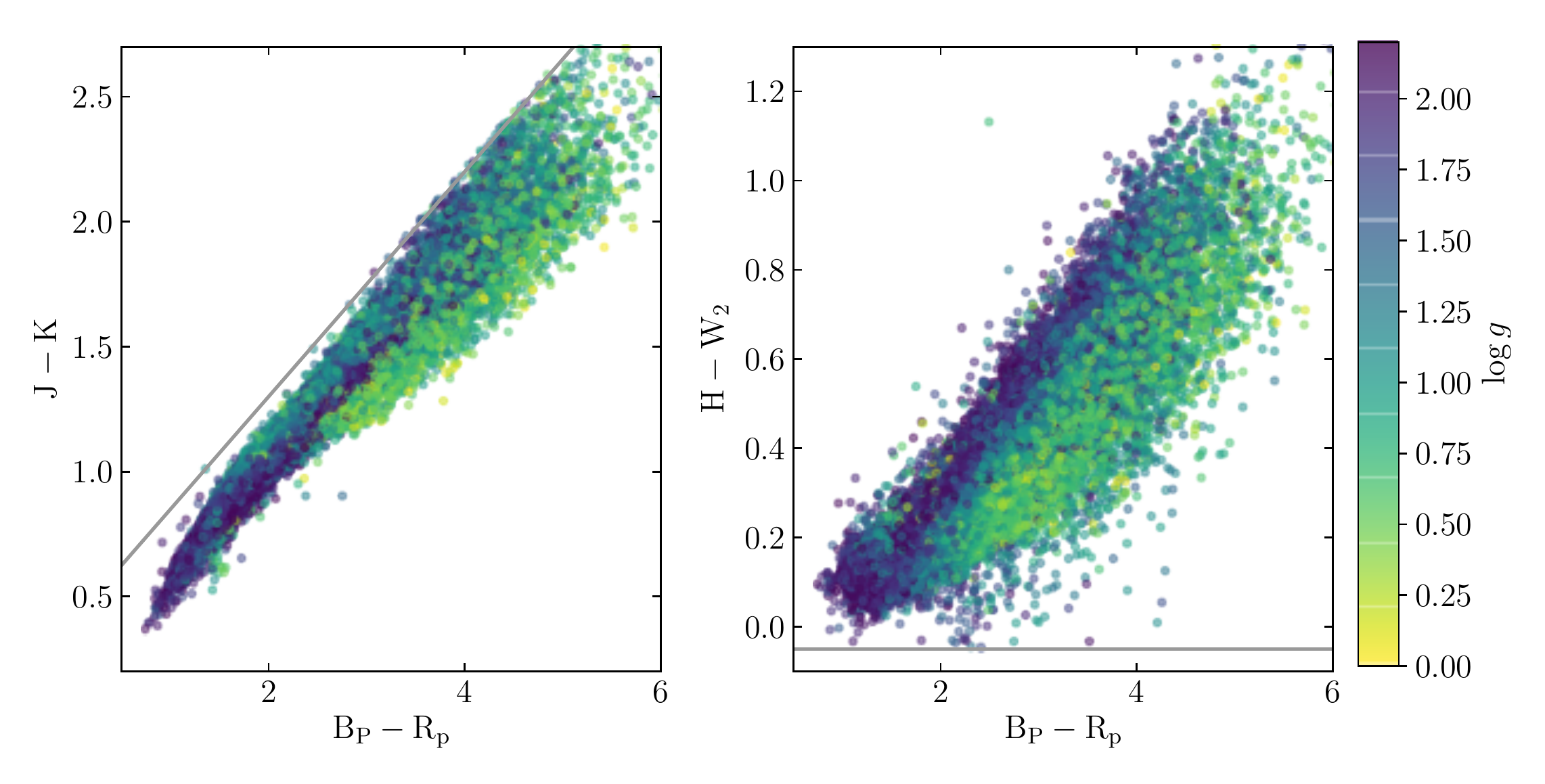}\\
\includegraphics[width=\textwidth]{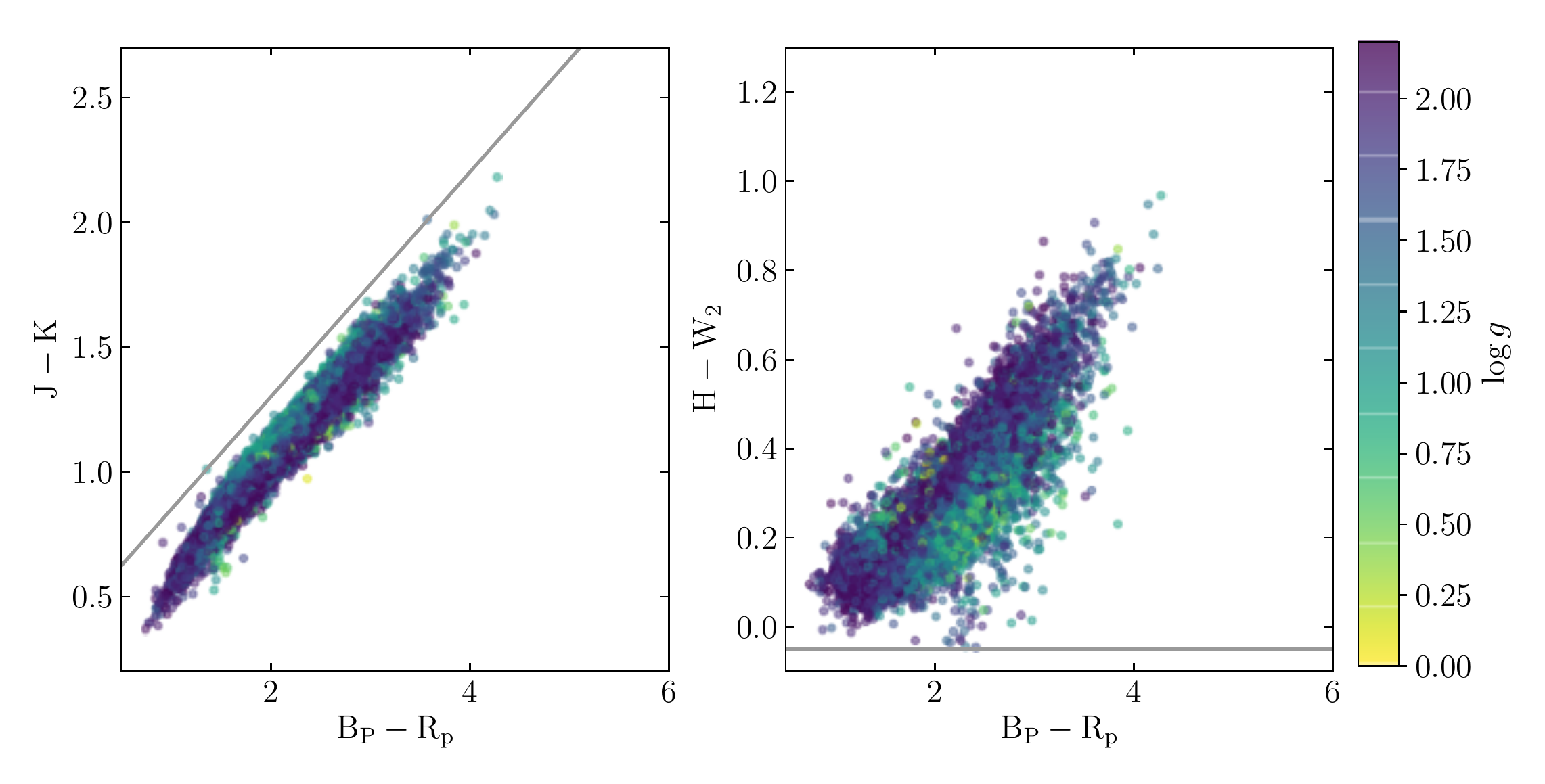}
\caption{\textsl{top:}
  Color--color diagrams for the 44784 Parent Sample stars.
  The points are colored by
  the \apogee\ pipeline estimates of surface gravity.
  The color cuts used to trim outliers are shown as grey lines.
  \textsl{bottom:}
  The same but for the 28226 Training Set stars.
  Note that there is less color range in the Training Set than in the
  Parent Sample.\label{fig:samples}}
\end{figure}

We need complete feature-vector information for all stars.  For this
reason, we define the Parent Sample to be the set of all stars that
meet the \apogee\ and \gaia\ cuts and also
have the complete set of photometry: $G_n$, $\BP_n$, $\RP_n$, $J_n$,
$H_n$, $K_n$, $W_{1n}$, and $W_{2n}$ and also an \apogee\ spectrum.
In addition, we applied two light color cuts to remove stars with obviously
contaminated or outlying photometry:
\begin{eqnarray}
(J_n - K_n) &<& (+0.4\,\mathrm{mag}) + 0.45 * (\BP_n - \RP_n)
\\
(H_n - W_{2n}) &>& (-0.05\,\mathrm{mag})
\quad .
\end{eqnarray}
These cuts removed roughly 2~percent of the \apogee\ stars.
This Parent Sample contains 44784 stars and is shown in \figurename~\ref{fig:samples}.

Every Parent Sample star gets, in addition, a randomly assigned binary
label (A or B).
This is used for two-fold validation and jackknife.
In short, we will
use the model trained on the Training Set A data to assign spectrophotometric parallaxes
to the Parent Sample B data and
use the model trained on the Training Set B data to assign spectrophotometric parallaxes
to the Parent Sample A data.

For training, we need A and B Training Sets.
We define the Training Set stars to be Parent Sample stars that also
have a measured \gaia\ parallax.
The Training Set stars, in addition to meeting all Parent Sample cuts,
also must meet an uncertainty criterion $\gsigma_n < 0.1\,\mathrm{mas}$, i.e. 
\begin{equation}
\code{parallax_error < 0.1}
\end{equation}
it must be observed by more than $8$ (widely separated) times
\begin{equation}
\code{visibility_periods_used >= 8}
\end{equation}
as well as meet the additional quality criterion
\begin{equation}
\code{astrometric_chi2_al / sqrt(astrometric_n_good_obs_al - 5) <= 35}
\end{equation}
which eliminates detections that are not consistent with a single PSF \citep[see][]{calj},
to ensure that the parallax measurements are good.
These cuts leave 28226 stars total in the Training Sets.
But---as we have emphasized above---we do not cut ever on parallax or
or the ratio to its uncertainty. For this reason, most of the Training Set stars
do not have significantly measured parallaxes and more than 2~percent even have
negative parallaxes!
But we will perform our training such that it will be unbiased in these
circumstances.
Indeed it would be necessarily biased if ever we did cut on parallax or
parallax signal-to-noise.

In detail,
for each star $n$ in the Training Sets, we take from \gaia\ \acronym{DR2} (\citealt{gaiadr2})
the astrometric parallax $\gparallax_n$ and its uncertainty $\gsigma_n$.
Importantly, to each \gaia\ parallax measurement we add a positive
offset $\varpi_0$ to adjust for the under-estimation of
parallaxes reported by the \gaia\ Collaboration (\citealt{lindegren}).
The \gaia\ recommended all-sky-average value is $\varpi_0=0.029$\,mas (\citealt{lindegren}),
but we adopt $\varpi_0 = 0.048$\,mas because that value
optimizes our cross-validation objective.
It also happens to be in the range of other values found in the literature (\citealt{arenou, zinn}).
Of course we really expect the offset to be a function of sky position and color (at least),
so the offset we find is not recommended for further use as any kind of universal value.
The spectrophotometric parallaxes we generate
do not depend strongly on this choice, except
for small differences at very small parallax (large distance).
The Training Sets are shown in the bottom panels of \figurename~\ref{fig:samples}.

Comparison of the top and bottom panels of {\figurename}~\ref{fig:samples} shows
that there is more color range in the Parent Sample than in the Training Set.
This challenges the representativeness assumption in \sectionname~\ref{sec:assumptions}.
The principal reason for the difference is that the \gaia\ quality cuts
exclude stars preferentially from crowded regions, which also tend to be the
most dust-obscured.
The hope of the model assumptions is that the Training Set will contain sufficient
dust variation that the model will naturally learn the dust corrections and extrapolate
acceptably.

For every star $n$ in the full Parent Sample we construct the feature
vector $x_n$ as
\begin{eqnarray}
x_n\T &\equiv& [1, G_n, \BP_n, \RP_n, J_n, H_n, K_n, W_{1n}, W_{2n}, \ln f_{1n}, \ln f_{2n}, \cdots, \ln f_{Ln}]
\end{eqnarray}
where the 1 permits a linear offset,
the photometry is from \gaia, \zmass, and \wise, respectively,
the fluxes in the $L=7405$ \apogee\ spectral pixels (for which there are reliably
and consistently data) for star $n$ are denoted
$f_{1n}$, $f_{2n}$, and so on,
and we have taken logarithms of those fluxes.
These feature vectors live in a $D$-dimensional space where $D=7414$.
For every star $n$ in either of the Training Sets we additionally require---in
addition to these feature vectors---a \gaia-measured astrometric parallax $\gparallax_n$
and uncertainty $\gsigma_n$.

\section{Results and validation}

\begin{figure}
\includegraphics[width=\textwidth]{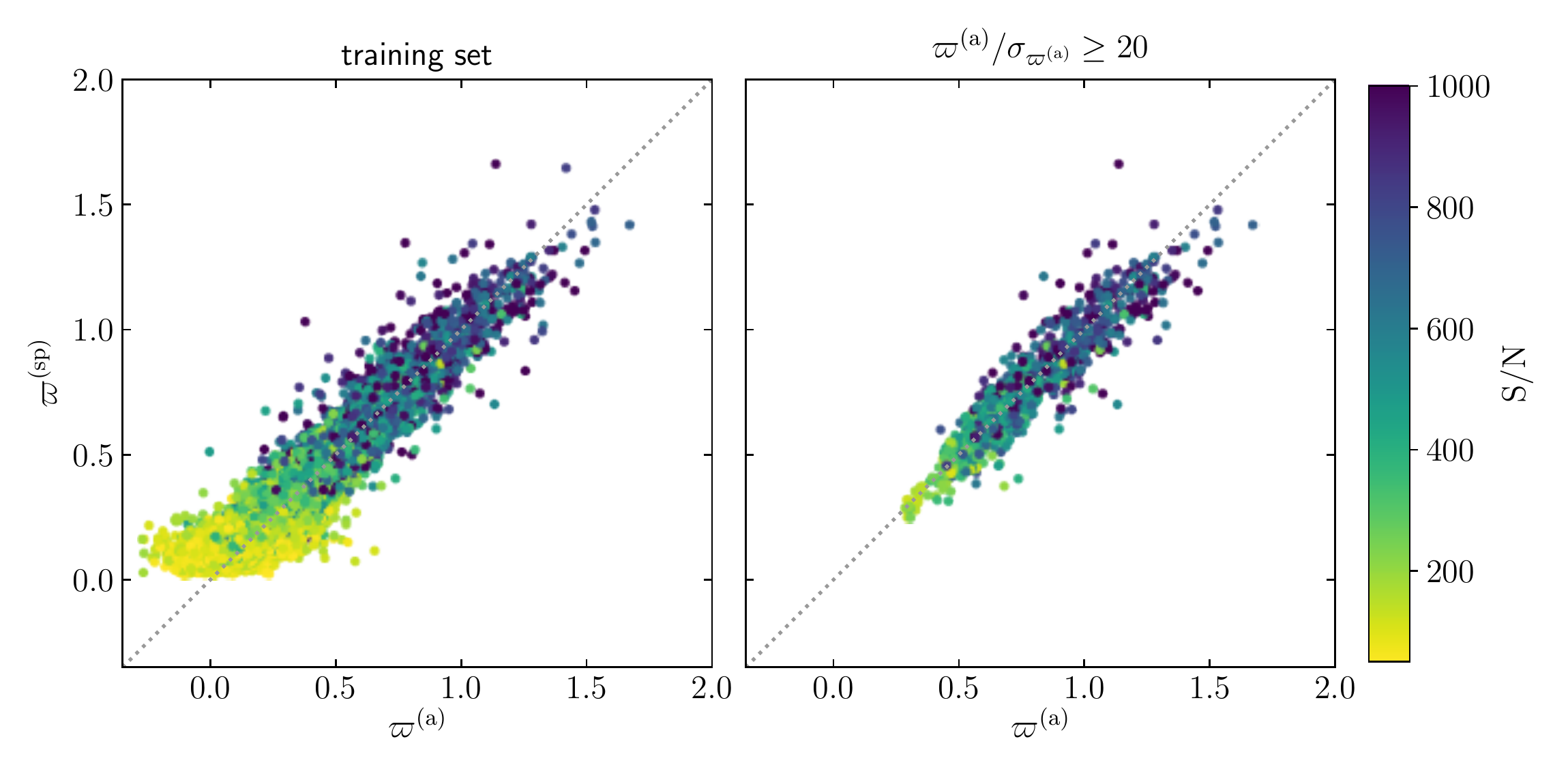}
\caption{The predictive accuracy of the model, in the two-fold cross-validation.
  Each panel shows \gaia\ astrometric parallaxes $\gparallax_n$ on the vertical axis
  and our spectrophotometric parallaxes $\sparallax$ on the horizontal axis.
  The left panel shows the full Training Set (both A and B combined),
  and the right panel shows a
  subset of stars with very high astrometric parallax signal-to-noise. This latter
  set plays no particular role in the method, but it can be used to demonstrate or
  assess the prediction precision. The fractional precision of the prediction in the
  right panel is better than 9 percent. Note that by construction, there can be no negative
  spectrophotometric parallaxes; that is, the quantity plotted on the vertical axis is
  positive-definite.\label{fig:xval}}
\end{figure}

We optimize two models, one for Training Set A and one for Training Set B.
The two models are applied to the two splits of the Parent Sample, Parent Sample
A and Parent Sample B, using the A-trained model on Parent Sample B and
\foreign{vice versa}.
For the purposes of assessing the accuracy and precision of the model, this
train-and-test framework constitutes a two-fold cross-validation.
Results of this cross-validation are shown in \figurename~\ref{fig:xval}.
For the stars with highest \gaia-measured parallax signal-to-noise,
the spectrophotometric model is predicting the \gaia\ parallaxes with little bias
and a scatter of less than 9~percent.

We used this cross-validation framework to adjust the regularization parameter $\Lambda$.
A coarse grid search in the value for the spectral-component value of the
$\lambda$ vector, using the A/B split as the two-fold cross-validation, with the objective
of the cross validation being prediction accuracy.
We get a prediction accuracy of better than 9 percent 
at the optimal setting of the regularization strength.

\begin{figure}
\includegraphics[width=\textwidth]{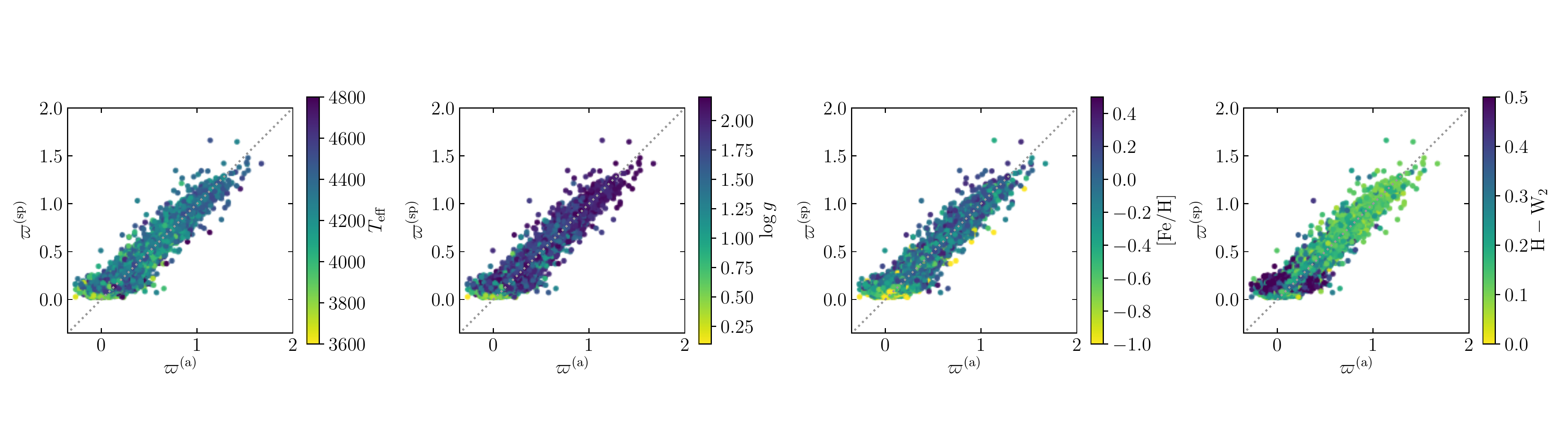}
\caption{Prediction validation colored by relevant features.
  Each panel shows the predicted parallax \foreign{vs} the
  spectrophotometric parallax colored by a different
  data feature that might be relevant to the residual.
  Again the quantity plotted on vertical axis is positive-definite.\label{fig:residuals}}
\end{figure}

This 9-percent precision estimate is conservative in the sense that it does not
deconvolve or correct for the contributions from the \gaia\ noise.
However, this test is performed with
\gaia's best stars, which are bright, nearby, in low extinction regions, and
(because of these things) near Solar metallicity.
There could in principle be additional bias or scatter for stars that are in dustier
regions or at lower metallicities.
\figurename~\ref{fig:residuals} shows that
there is no suggestion of any such trends when we color the residuals by metallicity
or $H-W_2$ color (which is a reddening proxy).
However, that figure does show that stars with greater reddening do appear to have
larger scatter against the \gaia\ parallaxes.
Additional tests confirm that our spectrophotometric parallaxes have largest scatter
in the dustiest and most crowded regions, as expected given the relatively low angular resolution
photometry.

\begin{figure}
~\hfill\includegraphics[width=0.7\textwidth]{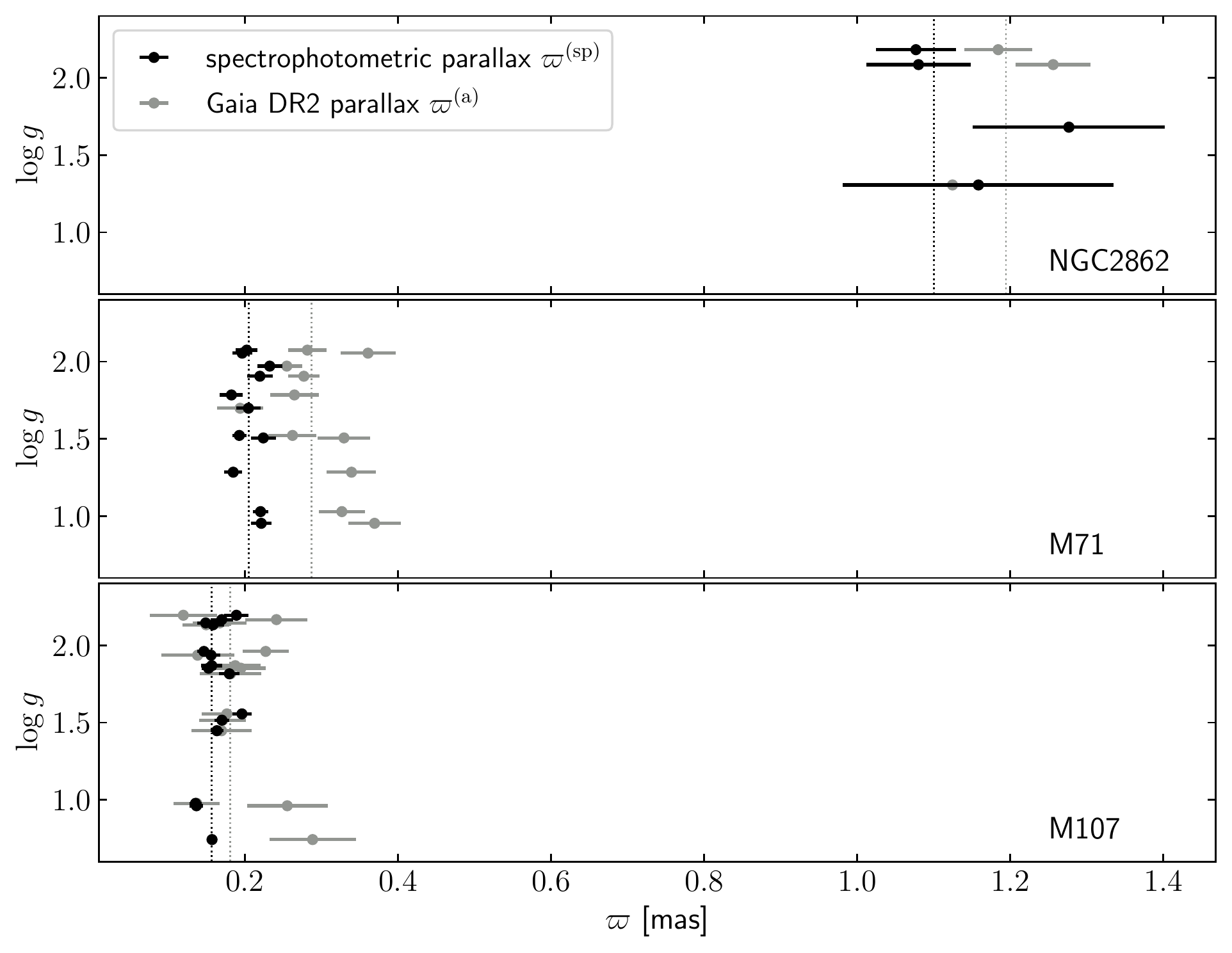}\hfill~
\caption{The \gaia\ astrometric parallaxes $\gparallax_n$
  and spectrophotometric parallaxes $\sparallax_n$ for member stars for
  three of the stellar clusters in the \apogee\ observing footprint. The spectrophotometric
  parallaxes and their inverse-variance-weighted mean and the astrometric parallaxes
  and their inverse-variance-weighted mean
  are shown as vertical lines.
  Cluster members were identified to be close to the cluster in celestial coordinates
  and comoving in proper motion.\label{fig:clusters}}
\end{figure}

Another validation of the results can be obtained by looking at known
clusters or spatially compact objects in the data.
In \figurename~\ref{fig:clusters}, we show the parallax distribution from \gaia\ and
the spectrophotometric-parallax distribution from our model for stars that are
astrometrically confirmed members of three of the stellar clusters in which we
validated our results.
The cluster members were found by hand, looking at sky and proper-motion distributions
centered on literature values.
In each case we are plotting only very securely identified cluster members; that
is, membership is conservative.
We chose the three displayed clusters
to span some range in metallicity and age and extinction, and therefore test
the accuracy of the method for different kinds of populations.
With the exception of \acronym{M71}, the clusters we tested
indicate that both \gaia\ and the spectrophotometric parallaxes appear to be unbiased
for these clusters and the range in abundances and ages they represent.
The case of cluster \acronym{M71} is troubling, but after searching for trends with
housekeeping data (also see \figurename~\ref{fig:residuals}) we don't find any correlations
with parallax offsets, although \acronym{M71} is a higher-extinction cluster.

\begin{figure}
\centering
~\hfill\includegraphics[width=0.8\textwidth]{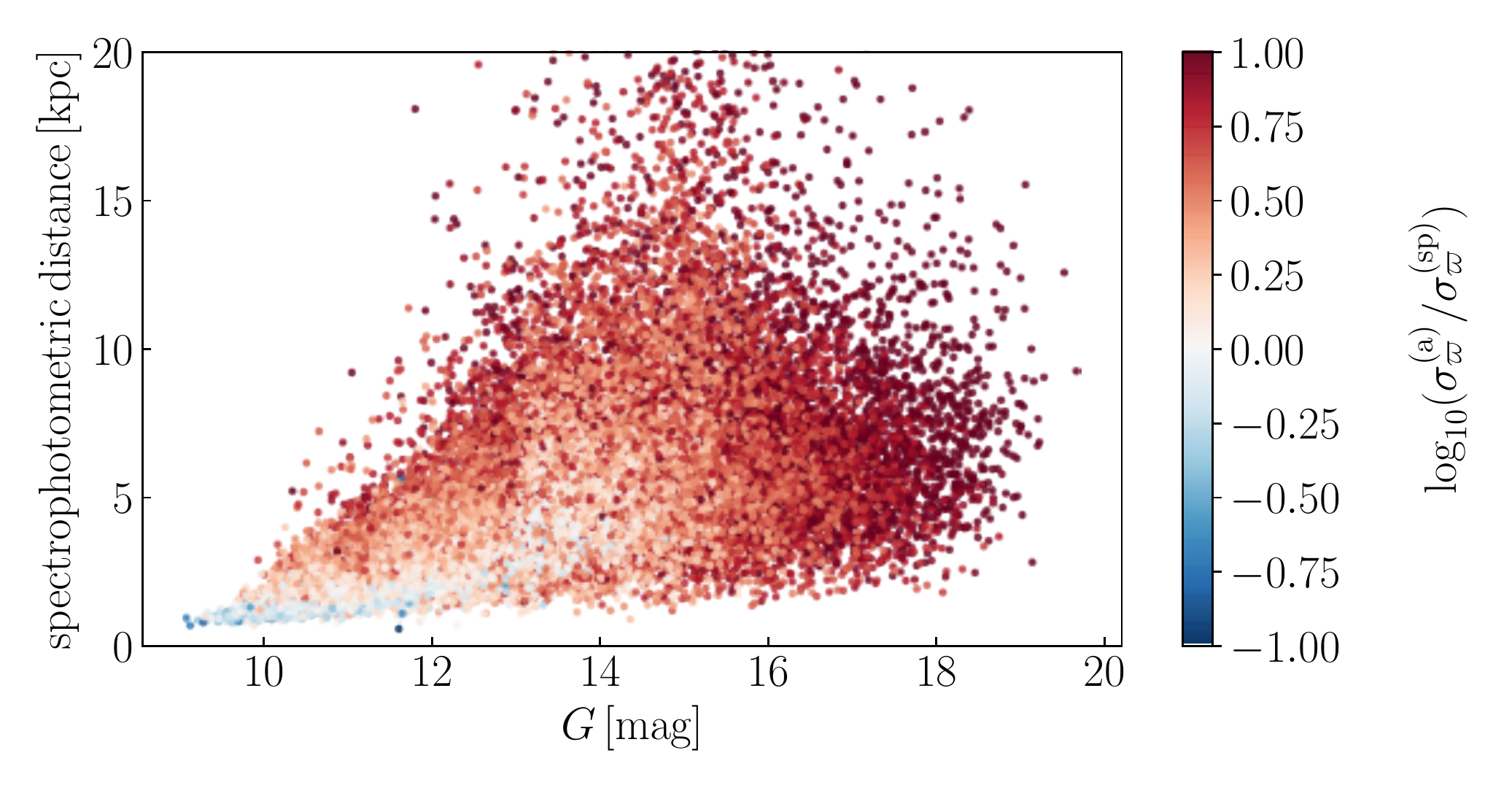}\hfill~
\caption{Comparison of precisions between the spectrophotometric parallaxes and
the photometric parallaxes, as a function of magnitude and heliocentric distance.
Note that the spectrophotometric estimates are more precise for most stars.
The information content of a parallax estimate (or any measurement)
is proportional to the inverse-square of the uncertainty.\label{fig:precision}}
\end{figure}

It is interesting and valuable to compare the spectrophotometric parallax precision to
the astrometric parallax precision.
\figurename~\ref{fig:precision} compares uncertainties between the spectrophotometric
outputs and the astrometric inputs
(and recall that information is proportional to the inverse square of the uncertainty).
The brief summary is that the spectrophotometric parallaxes are more precise (and even
far more precise) than the astrometric parallaxes for stars further than a few kpc from
the Sun.
This is not unexpected; \gaia\ parallax uncertainties are in the 0.1-mas range, whereas
our uncertainties are in the ten-percent range.

In the end, this method produces a spectrophotometric parallax estimate $\sparallax_m$
and uncertainty $\ssigma_m$ for every star $m$.
\tablename~\ref{tab:output} shows a snippet of this catalog of output, with the full catalog available
as an electronic data table associated with this \documentname.\footnote{\url{https://zenodo.org/record/1468053}}
In order to use this output with \gaia\ astrometry, we recommend joining our catalog to the
\gaia\ Catalog on the \zmass\ ID (name), as per the instructions at the \gaia\ Archive (\citealt{gaiaarchive}).
In order to use this output with \apogee\ radial velocities and element abundances, we recommend
joining our catalog to the \sdssiv\ Catalog on the \zmass\ ID (name).
In addition to this catalog, all the code used for this project is available online in an associated
code repository.\footnote{\url{https://github.com/aceilers/spectroscopic_parallax}}

\section{Discussion}

We have demonstrated that photometry and spectroscopy can be used to predict
or estimate distances to stars.
This isn't new!
However, what's new is that it is possible to do this very well for luminous giants
with unknown ages, and with a purely
linear model acting on magnitudes and spectral (logarithmic) fluxes, and entirely
data-driven.
That is, a linear model trained on noisy parallax data from \gaia.

The model and method are based on a raft of assumptions, listed above
in \sectionname~\ref{sec:assumptions}.
We make some comments there, but we emphasize some particulars here:

One strong assumption of this work is that our set of features is sufficient to
build a predictive model for distance.
Our argument for our included features is that they ought to be sufficient to
estimate a dust-independent apparent magnitude and the stellar parameters (especially
age and surface gravity) that are sensitive to luminosity.
In particular, since our model is linear in the logarithm of the parallax, it is
also linear in any kind of apparent magnitude at fixed distance and also linear
in distance modulus at fixed luminosity.
Our feature decisions were highly motivated by these kinds of considerations.

However, our choice of features is very rigid:
In principle the data themselves should tell us what features to include.
In that direction, it would make sense to choose features using a more complex
technique like deep learning or an auto-encoder.
These methods generate good features automatically.
They also involve an enormous set of hyper-parameter-like choices,
and they involve sacrificing certain kinds of interpretability.
All that said, we expect that a better set of features would do better; in this sense
the project here is just a first step towards precise spectrophotometric distance
estimates.

As we said above,
we required that the parallax prediction be constructed from a linear combination
of the input data or feature vector components.
This may seem like an absurd simplification; it begs the questions:
Why did it work? And could we have done better?
The answer to the first is that we have taken such a small part of stellar parameter
space with our surface-gravity cut, that the linearized model of stellar
spectrophotometry is not a bad approximation.
Also, because our linear model included photometry in magnitudes (that is, logarithmic)
and was exponentiated to make the parallax prediction, we knew in advance that much
of the needed flexibility (for dust attenuation, for example) would be close to
linear.
That is, it was a combination of limited model scope and some cleverness in the
feature engineering and model structure.

The answer to the second question is that almost certainly we could have done better!
Since models like Gaussian processes and deep networks subsume the linear model
(at least approximately), they could have delivered better---or at least non-worse---results.
The issues with going to a much more complex model are manifold, however:
We would have had many more decisions to make and many more hyper-parameters
to set.
We would have used some (or maybe much) of the information in the data to learn
the simple fact that the problem is close to linear; that is, if you give a model
a lot of freedom or capacity, it has to use a lot of the data to learn what part
of that capacity it really needs.
A more complex model might have had edge issues: Very flexible models don't extrapolate
outside the convex hull of their training data, and our training data was not
balanced in terms of parallax and dust attenuation.
In the  linear model it is also---unlike in more flexible models---trivial to propagate
uncertainties from the feature space to the prediction.
We used that, above, to propagate uncertainty in \equationname~(\ref{eq:unc}).
And finally---in principle---linear models are better for visualization and interpretation;
the linear model contains essentially only first derivatives in the data space, which are
relatively easy to understand.

Another  advantage of a linear model over more general methods (like deep learning, say),
is that it is possible to look inside
the linear model and check whether the dependencies represented there make sense.
For example, the feature vector $x_n$ for star $n$
contains a set of photometry in magnitudes.
Linear combinations of these photometric measurements are like complex synthetic
magnitudes or colors.
Similarly, linear combinations of spectral pixels are like a projection of the spectral space.
Brief inspection of these linear combinations showed them to be sensible, but it is out of
scope here to interpret them in detail.

Although the model, training, and prediction make no use whatsoever of stellar
models, we did use stellar models indirectly to select the Parent Sample of stars:
We used the \apogee\ pipeline (\citealt{aspcap}) surface-gravity $\logg$ measurements.
This tiny use of stellar models could have been obviated by selecting stars not
on the basis of the derived physical parameter, but instead selecting stars to
be similar, observationally, in the spectral space.
That would have lead to a clean sample and made the project independent completely
of stellar models.

Our uncertainty estimation is very naive; it presumes that the features are measured
with no intrinsic correlations (probably not true for neighboring pixels in the \apogee\ spectra,
for example) and it does not allow for any biases in the feature inputs.
What's shown in \figurename~\ref{fig:residuals} is that the biggest deviations are for
stars with large (red) $H-W_2$ colors.
These are in dusty and crowded regions; we expect that is the crowding that's causing the
greatest problem, because both \gaia\ and \wise\ (and probably also \zmass) photometry is
affected by overlapping and blended sources.

We validated the uncertainty estimates by looking at the chi-squared statistic and robust
variations of it.
We find that chi-squared is larger than expected if the errors are treated naively.
However, robust estimates show that this is driven by outliers;
median absolute differences are consistent with our expectations under this very naive noise model.
We also find that the outliers that drive up the chi-squared statistic tend to be red
in $H-W_2$.

\figurename~\ref{fig:clusters} suggests that there may be some hidden biases in
the results; the cluster M71 looks inconsistent between the \gaia\ and the spectrophotometric
parallaxes.
We have looked at colors, sky position, and cluster properties, and we don't have any
simple explanation for the discrepancy.
This is left as a caveat for the reader and our users.

Because of our train-and-test framework, each star is given a spectrophotometric
parallax based on coefficients derived from training on a complementary Training Set sample, of
which that star is not a member.
This framework ensures that the \gaia\ parallax $\gparallax_m$ of star $m$ is never used,
even indirectly, in generating the spectrophotometric parallax $\sparallax_m$ of star $m$.
That means that the parallax estimates generated here and given in \tablename~\ref{tab:output}
are statistically independent of the \gaia\ parallaxes and can be combined with them in
na\"ive ways, for example in inverse-variance-weighted averages.

\begin{figure}
\includegraphics[width=\textwidth]{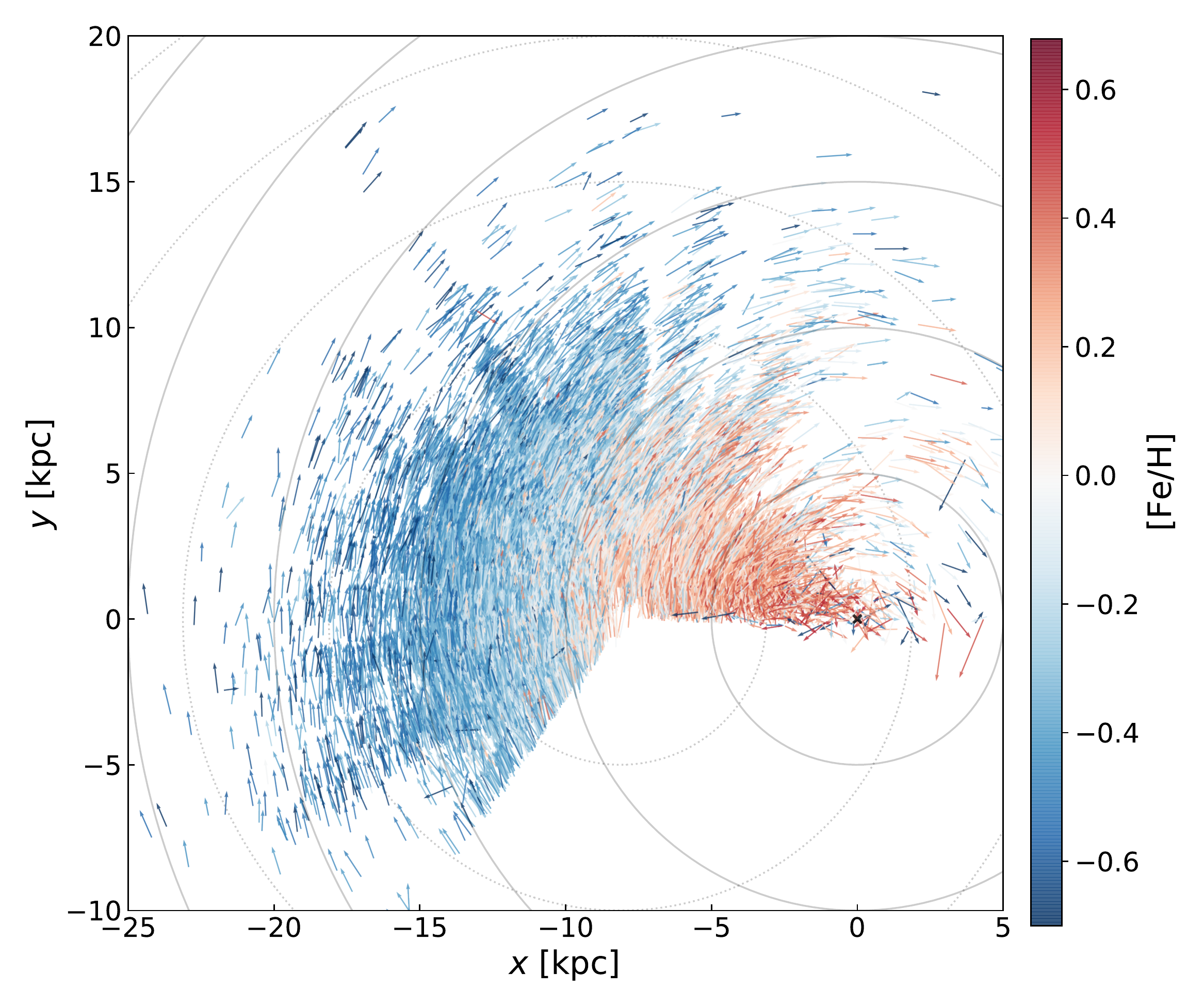}
\caption{A map of the kinematics and element abundances in the Milky Way disk.
  Each arrow represents the three-space Galactocentric velocity of each star
  in the sample, projected onto the Galactic disk plane.
  To make these velocities, raw \gaia\ proper motions and \apogee\ radial velocities
  were used (in a standard Galactocentric transformation), so they are
  not in any way noise-deconvolved.
  Each arrow has a color corresponding to the \apogee\ reported metallicity.
  This map shows the rotation of the Galaxy out to large radius, and also
  indicates the kinematic center of the disk rotation. The stars shown in this
  map have distance uncertainties on the order of ten percent, so at significant
  heliocentric distances, the interpretation of
  the quantitative information in this plot requires care.\label{fig:disk}}
\end{figure}

This \documentname\ and project delivers greatly improved distance estimates for
luminous red-giant stars, more luminous than the red clump.
The purpose of all this is to make global maps of the Milky Way.
Just as a demonstration, we show kinematics and element abundances for stars as a function
of position in the Milky Way disk in \figurename~\ref{fig:disk}.
In this visualization, it is possible to see the rotation of the disk and the
kinematic center of the Galaxy; that might even provide a distance estimate to the Galactic
Center.
It also shows extremely strong chemical gradients.
In a companion paper (\citealt{eilers}), we are using this information to estimate
the (asymmetric-drift-corrected) circular-velocity curve for the Milky Way.

Finally, one comment on the use of these spectrophotometric parallaxes.
They are noisy.
Precise use of these involves building a likelihood function and performing inference,
just as the precise use of the \gaia\ data presented here required the same.
We recommend making analogous inferences when using these outputs.

\acknowledgements
It is a pleasure to thank
  Coryn Bailer-Jones (\acronym{MPIA}),
  Sven Buder (\acronym{MPIA}),
  Andy L. Jones,
  Daniel Michalik (\acronym{ESTEC}),
  Melissa Ness (Columbia),
  Joel Zinn (OSU),
  and the participants in the \acronym{MPIA} Milky Way Group Meeting
  and the \apogee\ Science Telecon,
for help with this project.
This project was developed in part at the
2018 \acronym{NYC} Gaia Sprint, hosted by the Center for Computational Astrophysics of
the Flatiron Institute in New York City in 2018 June.

This work has made use of data from the European Space Agency (\acronym{ESA}) mission
\gaia\ (\url{https://www.cosmos.esa.int/gaia}), processed by the \gaia\ Data
Processing and Analysis Consortium (\acronym{DPAC},
\url{https://www.cosmos.esa.int/web/gaia/dpac/consortium}). Funding for the
\acronym{DPAC}
has been provided by national institutions, in particular the institutions
participating in the \gaia\ Multilateral Agreement.

This publication makes use of data products from the Two Micron All Sky Survey, which is a joint project of the University of Massachusetts and the Infrared Processing and Analysis Center/California Institute of Technology, funded by the National Aeronautics and Space Administration and the National Science Foundation.

This publication makes use of data products from the Wide-field Infrared Survey Explorer, which is a joint project of the University of California, Los Angeles, and the Jet Propulsion Laboratory/California Institute of Technology, funded by the National Aeronautics and Space Administration.

Funding for the \project{Sloan Digital Sky Survey IV} has been provided by the Alfred P. Sloan Foundation, the U.S Department of Energy Office of Science, and the Participating Institutions. \sdssiv\ acknowledges
support and resources from the Center for High-Performance Computing at
the University of Utah. The \project{\acronym{SDSS}} web site is \url{www.sdss.org}.

\sdssiv\ is managed by the Astrophysical Research Consortium for the 
Participating Institutions of the \project{\acronym{SDSS}} Collaboration including the 
Brazilian Participation Group, the Carnegie Institution for Science, 
Carnegie Mellon University, the Chilean Participation Group, the French Participation Group, Harvard-Smithsonian Center for Astrophysics, 
Instituto de Astrof\'isica de Canarias, The Johns Hopkins University, 
Kavli Institute for the Physics and Mathematics of the Universe / 
University of Tokyo, the Korean Participation Group, Lawrence Berkeley National Laboratory, 
Leibniz Institut f\"ur Astrophysik Potsdam,  
Max-Planck-Institut f\"ur Astronomie (Heidelberg), 
Max-Planck-Institut f\"ur Astrophysik (Garching), 
Max-Planck-Institut f\"ur Extraterrestrische Physik, 
National Astronomical Observatories of China, New Mexico State University, 
New York University, University of Notre Dame, 
Observat\'ario Nacional~/~\acronym{MCTI}, The Ohio State University, 
Pennsylvania State University, Shanghai Astronomical Observatory, 
United Kingdom Participation Group,
Universidad Nacional Aut\'onoma de M\'exico, University of Arizona, 
University of Colorado Boulder, University of Oxford, University of Portsmouth, 
University of Utah, University of Virginia, University of Washington, University of Wisconsin, 
Vanderbilt University, and Yale University.

\facilities{
\sdssiv(\apogee-2),
\gaia,
\zmass,
\wise}

\software{
\code{Astropy} \citep{astropy, astropy2},
\code{IPython} \citep{ipython},
\code{matplotlib} \citep{matplotlib},
\code{numpy} \citep{numpy},
\code{scipy} \citep{scipy}}

\begin{longrotatetable}
\startlongtable
\begin{deluxetable*}{ccccccc}
\tablecaption{The generated spectrophotometric parallaxes and their uncertainties..
This \tablename\ is just a snippet; the full catalog of all 44784 objects in the combined Parent Samples
is available online. In order to reproduce the {\figurename s} in this \documentname, this catalog
should be joined to the \gaia, \zmass, \wise\, and \apogee\ Catalogs by \zmass\ ID.\label{tab:output}}
\tablehead{\dcolhead{\texttt{2MASS\_ID}} & \dcolhead{\texttt{Gaia\_parallax}} & \dcolhead{\texttt{Gaia\_parallax\_err}} & \dcolhead{\texttt{spec\_parallax}} & \dcolhead{\texttt{spec\_parallax\_err}} & \dcolhead{\texttt{training\_set}} & \dcolhead{\texttt{sample}}}
\startdata
2M00000002+7417074 & 0.3191 & 0.0329 & 0.3185 & 0.0164 & 0 & B \\
2M00000317+5821383 & 0.3643 & 0.0510 & 0.4428 & 0.0183 & 0 & A \\
2M00000546+6152107 & 0.3154 & 0.0351 & 0.3292 & 0.0164 & 0 & B \\
2M00000797+6436119 & 0.0771 & 0.0841 & 0.1943 & 0.0194 & 1 & B \\
2M00001719+6221324 & 0.1395 & 0.0439 & 0.1564 & 0.0128 & 0 & B \\
2M00002005+5703467 & 0.3003 & 0.0247 & 0.2372 & 0.0203 & 0 & B \\
2M00002036+6153416 & 0.0657 & 0.0637 & 0.1347 & 0.0131 & 1 & A \\
2M00002118+6136420 & 0.1936 & 0.0323 & 0.2173 & 0.0133 & 1 & B \\
2M00002227+6223341 & 0.1965 & 0.0450 & 0.1461 & 0.0089 & 0 & B \\
2M00002472+5518473 & 0.0947 & 0.0207 & 0.1473 & 0.0096 & 1 & B \\
2M00002908+6140446 & 0.1675 & 0.0460 & 0.1914 & 0.0176 & 1 & A \\
2M00003061+5820590 & 0.1669 & 0.0339 & 0.3502 & 0.0290 & 1 & A \\
... & ... & ... & ... & ... & ... & ...  \\
 \enddata
\tablecomments{All floating-point numbers are angles in mas.}
\end{deluxetable*}
\end{longrotatetable}

\end{document}